\documentclass[aps,twocolumn,eqsecnum]{revtex4}

\usepackage{graphicx,float}

\begin{document}

\title{An effective field theory for coupled-channel scattering }

\author{Thomas D. Cohen }

\email{cohen@physics.umd.edu}

\affiliation{Department of Physics, University of Maryland,
College Park, MD 20742-4111}

\author{Boris A. Gelman}

\email{gelman@physics.arizona.edu}

\affiliation{Department of Physics, University of Arizona,
Tucson, AZ 85721}

\author{U. van Kolck}

\email{vankolck@physics.arizona.edu}

\affiliation{Department of Physics, University of Arizona,
Tucson, AZ 85721}

\affiliation{RIKEN BNL Research Center, Brookhaven National Laboratory,
Upton, NY 11973}

\begin{abstract}
The problem of describing low-energy two-body scattering for
systems with two open channels with different thresholds is
addressed in the context of an effective field theory.  In
particular, the problem where the threshold is unnaturally small
and the cross section at low energy is unnaturally large is
considered. It is shown that the lowest-order point coupling
associated with the mixing of the channels scales as
$\Lambda^{-2}$ rather than $\Lambda^{-1}$ (the scaling of the same-channel 
coupling and the scaling in a single-channel case) where
$\Lambda$ is the ultraviolet cutoff. The renormalization of the
theory at lowest order is given explicitly. The treatment of
higher orders is straightforward.
The potential implications for systems with deep open channels are discussed.
\end{abstract}



\maketitle
\section{Introduction}

During the past decade methods of effective field theory (EFT)
have been developed for problems of interest to low-energy nuclear
physics \cite{W}.  The underlying idea is based on the separation of
scales. If a problem has widely disparate scales and one is
interested in low-energy phenomena, then the details of the
short-distance physics are unimportant ---all that matters is its
effects on the long-range phenomena. Formally, EFT replaces these
short-distance interactions with contact interactions
containing an arbitrary number of derivatives. 
Power counting justifies truncating the
interaction Lagrangian and allows an {\it a priori} estimate of
errors at a given order. An archetypical example of a pure EFT
treatment of relevance in nuclear and molecular
physics is the EFT of short-range forces \cite{2Npless,3Npless,haloEFT}. 
This formalism is useful
when the low-energy two-body cross section is much larger than what is expected
based on the underlying scales. 
When ---as in the two-nucleon system---
there is a shallow real or virtual bound state,
there is also a
large scale separation between the scattering length and all other
scales in the problem. 
A power counting then requires that
the lowest-order contact interaction be iterated to all orders
while other effects can be treated perturbatively \cite{2Npless}. 
Ultimately, as
far as two-body scattering is concerned, this approach essentially
reproduces the effective range expansion \cite{Bethe}. 
The true
power of the method becomes clear when 
more bodies and/or
external probes ({\it e.g}, electroweak currents)
are coupled to the two-body system \cite{W}.
For example, recent success has been achieved in the 
systematic description of three-nucleon systems \cite{3Npless},
and promising avenues exist for the exploration of
many-nucleon systems (see, {\it e.g.}, Ref. \cite{manyNs}).

The purpose of the present paper is to develop an EFT treatment
for the nonrelativistic two-body scattering problems that have
more that one open channel with different thresholds. The
scattering in both channels can be treated as being 
unnaturally large at low energy.  
Our purpose in developing this expansion is partly
formal: the renormalization of such a problem is
characteristically different from that in the single-channel case,
and it is interesting to construct explicitly the renormalized
amplitudes. 
In addition, our analysis may be of
practical use in a few specialized cases of nucleon-nucleus
scattering.
Finally, it
provides some insight into EFT treatments of a very general class
of problems of importance to both nuclear and hadronic physics,
namely, the scattering of two bodies when there are open channels
even at zero kinetic energy.

\section{A coupled-channel scattering \label{cch} }

To set up the problem consider a case of a nucleon $N$ scattering
off of a nucleus $A$.  We assume that
the nucleus has a low-lying excited
state ---$A^*$--- with the excitation energy much smaller than a
typical scale of the interaction between $N$ and $A$. In addition,
the system has an unnaturally-large cross section at low energy.
The scattering reaction of a nucleon $N$ off a nucleus $A$ has
two channels:
\begin{eqnarray}
{\rm channel \,\, I:}    \,\,\,\,\,\,\,\,\,\, N\, A & \rightarrow
& N \,A
\nonumber \\
{\rm channel \, \, II:}    \,\,\,\,\,\,\,\,\,\, N\,A & \rightarrow
& N' \, A^* \label{channels}
\end{eqnarray}
Channel ${\rm I}$ corresponds to an elastic, and channel ${\rm II}$ to an
inelastic scattering. 
We assume for simplicity that the
scattering in both channels occurs in an $s$ wave in a singlet spin
and isospin state. 
However, 
we allow the outgoing nucleon in channel II to be in a different
state, in order to explicitly account for different charge states.

The formal development of this theory is designed to closely
follow the case of the pionless EFT and as such may be viewed as a
generalization of the effective range expansion. However, the
kinematics is quite different in the present case and the
generalized effective range expansion will be qualitatively
different from the usual one in the case of a single channel.
Formally, we develop an effective field theory which is valid for
the case where there are two unnaturally-light scales in the
problem. One of them is the threshold of the second
channel of the scattering reaction. This threshold
is given by the excitation energy of $A^*$,
\begin{equation}
\Delta=   m_{A^*}-m_{A} + m_{N'} -m_N\, ,\label{delta}
\end{equation}
where $m_A$, $m_{A^*}$, $m_N$ and $m_{N'}$ are the
masses of the nucleus $A$, its excited state $A^*$
and the two nucleon states, respectively. The
present treatment is applicable when
\begin{equation}
\Delta \ll {M^2 \over 2 \mu_1} \sim  {M^2 \over 2 \mu_2},
\label{cond}
\end{equation}
where $M\sim 1/R$ is the typical scale describing the underlying
interactions of short range $R$, and $\mu_{1}=m_N m_A /(m_N+m_A)$ and
$\mu_{2}=m_{N'} m_{A^*} /(m_{N'}+m_{A^*})$ are the reduced masses
for the $N\, A$ and $N'\, A^*$ systems, respectively.
We want to study the scattering at an energy $E$ comparable
to $\Delta$.
We take the scattering
cross section at such low energy to be very large compared to $4\,\pi\, R^2$. 
The second unnaturally small quantity is the
inverse scattering length, which generates the very large scattering
cross section.

As in a single-channel scattering, observables are obtained from
the on-shell $T$-matrix (or the physical scattering amplitude). 
The $T$-matrix for the scattering with two open channels has a form of a 
two-by-two square matrix. Diagonal entries of this matrix describe elastic
scattering while the two off-diagonal entries correspond to the
inelastic reactions. 
The scattering is considered here in the
center-of-mass frame of the initial $N\,A$ system.

In close analogy to single-channel scattering the
coupled-channel $T$-matrix is a solution of a
Lippmann-Schwinger-like equation,
\begin{widetext}
\begin{equation}
T_{ij}(\vec{p}, \vec{q})=-V_{ij}(\vec{p}, \vec{q})+ \int {d^3 k
\over (2\pi)^{3}} \, V_{ir}(\vec{p}, \vec{k}) \,
G_{rs}(|\vec{p}|, |\vec{k}|)\, T_{sj}(\vec{k}, \vec{q}) \, ,
\label{LS}
\end{equation}
where $\vec{p}$ and $\vec{q}$ are the relative momenta of the
initial and final particles in the center-of-mass frame. The
interaction potential $V_{ij}(\vec{p}, \vec{q})$, and the
coupled-channel two-body non-relativistic Green function are
two-by-two matrices. The latter is a diagonal matrix; neglecting
terms of higher order in momenta over large masses,
\begin{equation}
G(|\vec{p}|, |\vec{k}|) = \left(
\begin{array}{c c}
{2\, \mu_{1} \over \vec{p}^2-\vec{k}^2+i\epsilon}  & 0 \\
0 & {2\, \mu_{2} \over \vec{p}^2 - 2\mu_{2}\Delta-\vec{k}^2
+i\epsilon}
 \end{array} \right) \, ,
\label{G}
\end{equation}
where $\Delta$ is the kinematical threshold for the channel ${\rm II}$
defined in Eq.~(\ref{delta}). Note that in the present regime the threshold
$\Delta$ is
positive---channel ${\rm II}$ is energetically above channel
${\rm I}$. In section~\ref{optical} we will consider a case when
$\Delta$ is negative---channel ${\rm II}$ is below channel ${\rm I}$.

The diagonal elements of $G(|\vec{p}|, |\vec{k}|)$, $G_{11}$ and $G_{22}$, 
describe off-shell
propagation of $N\,A$ and $N'\,A^*$ systems, respectively. These
Green functions can be obtained from the non-interacting part of
the total Lagrange density,
\begin{eqnarray}
{\cal L}(x)&=& \Psi_{N}^{\dag} (x) \left(i\, \partial_0 +{\nabla^2
\over 2\,m_{N}}\right) \Psi_{N} (x) 
+ \Psi_{N'}^{\dag} (x) \left(i\, \partial_0 - (m_{N'}-m_N) +{\nabla^2
\over 2\,m_{N}}\right) \Psi_{N'} (x) \nonumber\\
&&+  \Psi_{A}^{\dag} (x)
\left(i\, \partial_0 +{\nabla^2
\over 2\,m_{A}}\right) \Psi_{A} (x) +
\Psi_{A^*}^{\dag} (x)
\left(i\,
\partial_0 -(m_{A^*}-m_A) + {\nabla^2 \over 2\,m_{A}}\right) \Psi_{A^*} (x)
+ \dots \nonumber\\
&&+ {\cal L}_{\rm int}(x)\, ,\label{L}
\end{eqnarray}
\end{widetext}
where the field $\Psi_{\alpha}^{\dag} (x)$ ($\Psi_{\alpha} (x)$)
creates (destroys) a particle of type $\alpha$.
The ellipsis in Eq.~(\ref{L}) stand for the relativistic corrections.
Eq.~(\ref{L}) gives Eq.~(\ref{G}) apart from higher-order terms,
since in leading order $\mu_2=\mu_1$.

To obtain the $T$-matrix from Eq.~(\ref{LS}) we need to know the
form of the coupled-channel interaction potential $V(\vec{p},
\vec{q})$. It describes the interactions between the incoming
nucleon $N$ and the nucleons inside the nucleus $A$. In general,
this is a many-body problem. However, as long as the kinetic
energy is small so that the wavelength of the nucleon $N$ is
much larger than the range of the interaction, the effects of
this interaction on the low-energy scattering observables can be
represented 
by contact interactions between a nucleon and
the nucleus as a whole. Since the threshold $\Delta$ is of order of
the kinetic energy $E$, each nuclear state has to be considered as an explicit
degree of freedom.

The effective potential, $V(\vec{p}, \vec{q})$, is given by an
infinite sum of contact interactions, which in momentum space
has the form
\begin{equation}
V_{ij}(\vec{p},
\vec{q})=C_{0}^{ij}+C_{2}^{ij}(\vec{p}^{\,2}+\vec{q}^{\,2}) + \dots
\label{V} \,,
\end{equation}
where ellipsis stand for terms with higher powers of momenta
\footnote{Only terms that contribute to $s$-wave scattering
are shown.}. The constant coefficients  $C_{0}^{ij}$,
$C_{2}^{ij}$, {\it etc.}, are two-by-two matrices. For example,
\begin{equation}
C_0 = \left(
\begin{array}{c c}
c_{11} & c_{12} \\ c_{12} & c_{22}
 \end{array} \right) \,.
\label{C0}
\end{equation}

The potential in Eq.~(\ref{V}) can be obtained from 
the most general Lagrangian satisfying Lorentz, parity
and time-reversal invariance. The interaction part of
the Lagrangian in Eq.~(\ref{L})
is
\begin{widetext}
\begin{equation}
{\cal L}_{\rm int}(x)=- c_{11}
\Psi_{N}^{\dag}\,\Psi_{A}^{\dag}\, \Psi_{N}\, \Psi_{A}\,-
c_{22} \Psi_{N'}^{\dag}\, \Psi_{A^*}^{\dag}\, \Psi_{N'}\,
\Psi_{A^*}\, - c_{12} \left( \Psi_{N}^{\dag}\,
\Psi_{A}^{\dag}\, \Psi_{N'}\, \Psi_{A^*}\, + {\rm h.c.} \right) +
\dots \, ,\label{Lint}
\end{equation}
\end{widetext}
where the last term represents the coupling between the two channels, and
the ellipsis stand for terms containing derivatives.

The existence of an infinite number of singular interactions in
Eq.~(\ref{V}) requires one to introduce a power counting,
which justifies a truncation in
the number of interaction terms that must be kept up to a given
order in an expansion in powers of $E/M$.
In addition, as in any field theory,
a renormalization scheme is necessary to relate 
parameters in the Lagrangian to observables, and
render the latter independent of the cutoff.

In the case of a single-channel scattering at low energies with
short-range forces, this situation has been thoroughly
investigated and a well-defined EFT exists \cite{2Npless,3Npless}.
The power-counting scheme
depends on 
whether the scattering length is natural or unnatural;
in the latter case, it
relies on the fact that other
scattering observables, such as effective range and shape terms in
the effective range expansion, are natural,
as one would naively expect
when the underlying interactions are short ranged. 
The large scattering length usually indicates the presence of a 
shallow real
or virtual bound state. In this case an $s$-wave $T$-matrix
at leading order is obtained by iterating (via the
Lippmann-Schwinger equation) the interaction potential that is
given by a momentum-independent contact interaction containing a
single constant ---$c_0$ \cite{2Npless}.
In this iteration,
any regularization and renormalization scheme can be used:
momentum cutoff, dimensional regularization with power-divergence
subtraction, {\it etc.} \cite{renormpless}. 

As it will become clear below, it is possible to define a power counting
in the coupled-channel case in a similar way. However, the power
counting in the present case has to reflect the existence of two
unnaturally small scales, namely the large scattering
cross sections and the small threshold $\Delta$.

In this work, the infinite momentum integrals appearing in
Eq.~(\ref{LS}) will be regularized using a sharp momentum cutoff
$\Lambda$; however, the final results are independent of the
regularization procedure. As in a single-channel case with
unnatural scattering length, we expect that at leading order the
coupled-channel $T$-matrix is obtained by iterating the first
term in Eq.~(\ref{V}) ---the constant matrix $C_0$. The rest of
the terms are expected to be treated perturbatively according to
the mass dimension of the constants $C_{2n}$. Note that while
such a power counting seems to be a quite logical generalization
of the power counting in the single-channel case, it is not
totally obvious how to implement the renormalization for the
present case. One principal result of this paper is to
demonstrate how this can be accomplished.

The success of the power counting for single-channel scattering
is ultimately connected to an
ability to ``fine tune'' one ``bare'' constant $c_0(\Lambda)$ such as to
reproduce the scattering
length $a_s$, which can be much
larger than the typical scale $R\sim 1/M$ of the underlying 
short-range interactions. 
The single-channel $s$-wave $T$-matrix at leading order
has the following form (in the center-of-mass frame of two
particles with reduced mass $\mu$ and the initial relative
momentum $\vec{p}$) \cite{2Npless}:
\begin{widetext}
\begin{equation}
{1\over T}= {\mu \over 2 \pi}\left\{-{2\pi\over \mu\,c_0(\Lambda)} - {2
\Lambda\over \pi} -i\,p + {\cal O} \left({p^2 \over \Lambda}
\right) \right\}= {\mu \over 2 \pi}\left\{-{1\over a_s}-i\,p +
{\cal O} \left({p^2 \over \Lambda} \right) \right\}. \label{Ts}
\end{equation}
\end{widetext}
The right-hand side of Eq.~(\ref{Ts})
corresponds to the effective range expansion in which only the
first term ---the scattering length--- is reproduced; interactions
with more derivatives account for the other effective range
parameters. As follows from
Eq.~(\ref{Ts}), in order to obtain a $\Lambda$-independent
on-shell $T$-matrix at leading order, 
the bare coupling constant $c_0$ has to have the
following dependence on the cutoff scale $\Lambda$,
\begin{equation}
{1\over c_0 (\Lambda)}= {\mu \over 2\pi}\left\{{1\over a_s} - {2\Lambda
\over \pi} \right\}\,. \label{cL}
\end{equation}
The above equation determines the renormalization-group flow of
the coupling $c_0$. 

In the present case of coupled-channel scattering, 
the leading-order interaction contains three bare coupling constants,
Eq.~(\ref{C0}). Since at this order the coupled-channel
potential $V(\vec{p}, \vec{q})$ (Eq.~(\ref{V}))
is momentum independent, the coupled-channel Lippmann-Schwinger equation, 
Eq.~(\ref{LS}), can be solved analytically to all orders in the coupling
constants $C_{0}^{ij}$. The solution is,
\begin{equation}
T_s =-\,(1-C_0 \, G^{\Lambda})^{-1}\, C_0 \,, \label{Tsolution}
\end{equation}
where the cutoff-dependent matrix $G^{\Lambda}$ is
\begin{widetext}
\begin{equation}
G^{\Lambda}= \left(
\begin{array}{c c}
-{\mu_{1} \, \Lambda \over \pi^2} - {\mu_1 \over 2\pi}\, i p & 0 \\
0 & - {\mu_{2} \, \Lambda \over \pi^2}- {\mu_2 \over 2\pi}\,
i \sqrt{\vec{p}^2 -2\mu_{2}\Delta}
\end{array} \right) \,. \label{GL}
\end{equation}
Note that, due to the particularly simple
form of the coupled-channel potential at leading order, 
the on-shell $s$-wave 
$T$-matrix (or the physical coupled-channel scattering amplitude) is a 
function of the energy $E=|\vec{p}|^2/2\mu_{1}$ only
(for a fixed threshold $\Delta$). 
As in the single-channel case, the $\Lambda$-dependent terms come from the 
linearly divergent part of integrals in Eq.~(\ref{LS}) regularized using 
a momentum cutoff $\Lambda$. 
We have already neglected terms that depend on inverse powers of
the cutoff and are analytic in $E$; these terms cannot
be considered separately from the higher-derivative terms in
Eq.~(\ref{V}), which are also analytic in $E$.
Using Eqs.~(\ref{C0}) and (\ref{GL}), the explicit form of the coupled-channel
$T$-matrix at leading order is
\begin{equation}
T_s=-\frac{1}{1-(c_{11}G^{\Lambda}_{11}+c_{22}G^{\Lambda}_{22})
               - (c_{12}^2-c_{11}c_{22}) G^{\Lambda}_{11} G^{\Lambda}_{22}} 
    \left(\begin{array}{c c}
    {c_{11} +(c_{12}^2-c_{11} c_{22}) G^{\Lambda}_{22} }& 
    {c_{12}} 
    \\ \\
    {c_{12}} &
    {c_{22} +(c_{12}^2-c_{11} c_{22}) G^{\Lambda}_{11}}
 \end{array} \right) \, . \label{TL}
\end{equation}
\end{widetext}

As written, the $T$-matrix in Eq.~(\ref{TL}) contains 
$\Lambda$-dependent terms. As in the single-channel case (Eq.~(\ref{Ts})),
one can expect to obtain cutoff independent $T$-matrix elements 
because the bare coupling constants $c_{ij}$ depend on $\Lambda$.
Note that in addition to linearly-divergent terms ($G^{\Lambda}_{11}$
and $G^{\Lambda}_{12}$ separately), 
the $T$-matrix elements in
Eq.~(\ref{TL}) contain a product that scales as
$\Lambda^2$ ---$G^{\Lambda}_{11} \, G^{\Lambda}_{22}$.
Moreover, the form in which this product appears is
non-trivial.

In order to determine the cutoff dependence of the coupling constants 
$c_{11}$, $c_{22}$ and $c_{12}$, it is convenient to consider the inverse of
the $T$-matrix given in Eq.~(\ref{TL}),
\begin{widetext}
\begin{equation}
T^{-1}_{s}=\left(
\begin{array}{c c}
{c_{22} \over c_{12}^2-c_{11} c_{22}} 
-{\mu_{1} \, \Lambda \over \pi^2} - {\mu_1 \over 2\pi}\, i p & 
- {c_{12} \over c_{12}^2-c_{11} c_{22}} \\ \\
- {c_{12} \over c_{12}^2-c_{11} c_{22}} &
{c_{11} \over c_{12}^2-c_{11} c_{22}} 
 - {\mu_{2} \, \Lambda \over \pi^2}- {\mu_2 \over 2\pi}\,
i \sqrt{\vec{p}^2 -2\mu_{2}\Delta}
\end{array} \right) \, , \label{TLinv}
\end{equation}
\end{widetext}
which is the coupled-channel analog of the first part of the 
equation~(\ref{Ts}). 

The requirement that the $T$-matrix be cutoff-independent (at a given order
in an EFT expansion) leads to 
\begin{eqnarray}  
{c_{22} \over c_{11} c_{22}-c_{12}^2} & + & {\mu_{1} \, \Lambda \over \pi^2}
= {\mu_1 \over 2\pi \,a_{11}} \nonumber \\
{c_{11} \over c_{11} c_{22}-c_{12}^2} & + & {\mu_{2} \, \Lambda \over \pi^2}
= {\mu_2 \over 2\pi \,a_{22}} \nonumber \\ 
{c_{12} \over c_{11} c_{22}-c_{12}^2} & = &
{\sqrt{\mu_{1} \mu_{2}} \over 2 \pi a_{12}} \, ,
\label{ccL}
\end{eqnarray}
where $a_{11}$, $a_{22}$ and $a_{12}$ are constants,
the constants of proportionality having been chosen for further convenience. 
The
system of coupled equations in Eq.~(\ref{ccL}) can be solved to obtain the
$\Lambda$-scaling of $c_{11}$, $c_{22}$ and $c_{12}$,
\begin{widetext}
\begin{eqnarray}
{1 \over c_{11} (\Lambda)} &  = & {\mu_1 \over 2\pi} \left\{{1 \over a_{11}}
-{2\Lambda \over \pi} - {1 \over a_{12}^{2}} \,
{1 \over {1 \over a_{22}} - {2 \Lambda \over \pi}} \right\}= 
{\mu_1 \over 2\pi} \left\{{1 \over a_{11}} -{2 \Lambda \over \pi} +
{\cal O}\left({1\over a_{12}^2\Lambda} \right) \right\} \, , \nonumber \\
{1\over c_{22} (\Lambda)} & = & {\mu_2 \over 2\pi} \left\{{1 \over a_{22}}
-{2\Lambda \over \pi} - {1 \over a_{12}^{2}} \,
{1 \over {1 \over a_{11}} - {2 \Lambda \over \pi}} \right\}=
{\mu_2 \over 2\pi} \left\{{1 \over a_{22}} -{2 \Lambda \over \pi} +
 {\cal O}\left({1\over a_{12}^2\Lambda} \right) \right\}\, , \nonumber \\
{1\over c_{12} (\Lambda)} & = & {\sqrt{\mu_1\,\mu_2}\over 2\pi} a_{12} 
\left\{ \left( {1 \over a_{11}}- {2\Lambda \over \pi} \right) 
        \left( {1 \over a_{22}}- {2\Lambda \over \pi} \right)
        - {1 \over a_{12}^{2}} \right\}\, .
\label{cii}
\end{eqnarray}
Note, the $\Lambda$-dependence of the coupling constants $c_{11}$
and $c_{22}$ is the same up to terms of order $1/\Lambda$ as 
the $\Lambda$-scaling of the coupling constant $c_0$ 
in the single-channel case,
Eq.~(\ref{cL}). The $\Lambda$-scaling of $c_{12}$, on the
other hand, is qualitatively
different: it scales as $\Lambda^{-2}$.

Using Eq.~(\ref{cii}) in Eq.~(\ref{TLinv}) 
we obtain the cutoff-independent form
for the inverse of the coupled-channel $T$-matrix,
\begin{equation}
T^{-1}_{s}=\left(
\begin{array}{c c}
- {\mu_1 \over 2\pi}\,\left( {1\over a_{11}}+ i p \right) & 
{\sqrt{\mu_{1} \mu_{2}} \over 2 \pi a_{12}}
\\ \\
{\sqrt{\mu_{1} \mu_{2}} \over 2 \pi a_{12}} & 
-{\mu_2 \over 2\pi}\, \left( {1\over a_{22}} +
i \sqrt{\vec{p}^2 -2\mu_{2}\Delta} \right)
\end{array} \right) \, , \label{Tsinvlimit}
\end{equation}
which depends only on the physical constants $a_{11}$, $a_{22}$ and $a_{12}$.
Thus, the coupled-channel $T$-matrix at leading order is,
\begin{equation}
T_{s}   =
  \left\{ {1\over a_{12}^{2}}- \left({1\over
a_{11}}+i\,p \right) \left({1\over a_{22}}+i\,\sqrt{\vec{p}^2
-2\mu_{2}\Delta}\right) \right\}^{-1}
  \left(
\begin{array}{c c}
 {2\pi\over \mu_1} ({1\over a_{22}}+i\,\sqrt{\vec{p}^2 - 2\mu_{2}\Delta}) &
 {2\pi\over \sqrt{\mu_1\,\mu_2}} \, {1\over a_{12}} \\ \\
 {2\pi\over \sqrt{\mu_1\,\mu_2}} \, {1\over a_{12}} &
 {2\pi\over \mu_2} ({1\over a_{11}}+i\,p)
 \end{array} \right) \, ,  \label{TT}
\end{equation}
\end{widetext}
where $p=\sqrt{2\mu_{1} E}$ with $E$ being the center-of-mass
energy in channel ${\rm I}$. 

Equation~(\ref{TT}) represents the leading term in the
generalized effective range expansion of the coupled-channel
$s$-wave $T$-matrix for a system with short-range interactions. 
It
depends on three constants ---$a_{11}$, $a_{22}$ and $a_{12}$---
that play the same role as the scattering length in the single-channel
case. Higher-order terms can be obtained in similar fashion.
Likewise, more channels can be included.
We see that the EFT reproduces the standard multi-channel
generalization of the effective range expansion \cite{ccERE}.
As expected, the existence of the low-energy
threshold (given by $\Delta$) for the second reaction channel
(channel ${\rm II}$) modifies the scattering in the initial
channel ---channel ${\rm I}$.
Even though Eq.~(\ref{Tsinvlimit}) is an obvious generalization 
of the single-channel case, it has some noteworthy features.

When the kinetic energy $E < \Delta$, channel ${\rm II}$ is closed.
Thus, we expect that the expansion of $T_{s11}$ (describing the
scattering in channel ${\rm I}$) in powers of the relative momentum
$\vec{p}$ takes the same form as the effective range expansion in
the single-channel case, Eq.~(\ref{Ts}). Indeed, expanding the inverse
of the element $T_{s11}$ in Eq.~(\ref{TT}) in powers of
$p^2/\mu_2 \Delta$, one obtains
\begin{widetext}
\begin{equation}
{1\over T_{s11}} ={\mu_1 \over 2\pi} \left\{-{1\over a_{\rm eff}}
+ {1\over 2}\, r_{\rm eff} \, p^2 -i\, p 
+{\cal O}\left({p^4\over a_{12}^2 (\mu_2 \Delta)^{5/2}}\right) \right\} \, , 
\label{efr1}
\end{equation}
\end{widetext}
where the effective range parameters are
\begin{eqnarray}
a_{\rm eff} & \equiv &  {a_{11} a_{12}^{2}(1- a_{22} \sqrt{2
\mu_{2} \Delta}) \over a_{12}^2(1-a_{22} \sqrt{2 \mu_{2}
\Delta})-a_{11} a_{22}} \; , \nonumber \\
r_{\rm eff} & \equiv &  -{1\over \sqrt{2 \mu_{2}\Delta}}
\left( {a_{22} \over a_{12} (1-a_{22} \sqrt{2 \mu_{2}\Delta})}\right)^2
\,. 
\end{eqnarray}

The above expansion has indeed the form of the effective range expansion. 
Despite channel ${\rm II}$ being closed ($E <\Delta$), 
the fact that the threshold $\Delta$ is much smaller
than the scale $M$ of the short-range interactions changes the
scaling of the effective range parameters. 
While the
effective range for the single-channel case ---or, effectively,
for coupled channels 
when $\sqrt{\mu_2 \Delta} \sim M$--- 
is of order $1/M$, 
here it is of order $1/\sqrt{\mu_{2} \Delta} \gg 1/M$.
The same is true for other effective range parameters.
{}From the perspective of channel ${\rm I}$ alone,
all effective range parameters are large.

The off-diagonal elements of the coupled-channel $T$-matrix,
Eq.~(\ref{TT}), are not identically zero
even for the case when $E < \Delta$. This, however, does not mean
that there is a non-zero cross section for the scattering into
channel ${\rm II}$. The reason is that the asymptotic wave function
in channel ${\rm II}$ is, up to a constant, 
$$ {1\over r} {\rm e}^{-i\, r \, \sqrt{p^2- 2 \mu_{2} \Delta}}, $$
which gives zero contribution to the outgoing flux when $p^2 < 2
\mu_{2} \Delta$ and, as a result, the scattering into this channel
has zero cross section.

\begin{figure}[t]
\begin{minipage}[t]{1\columnwidth}
\centering
\includegraphics[height=7cm,width=9cm]{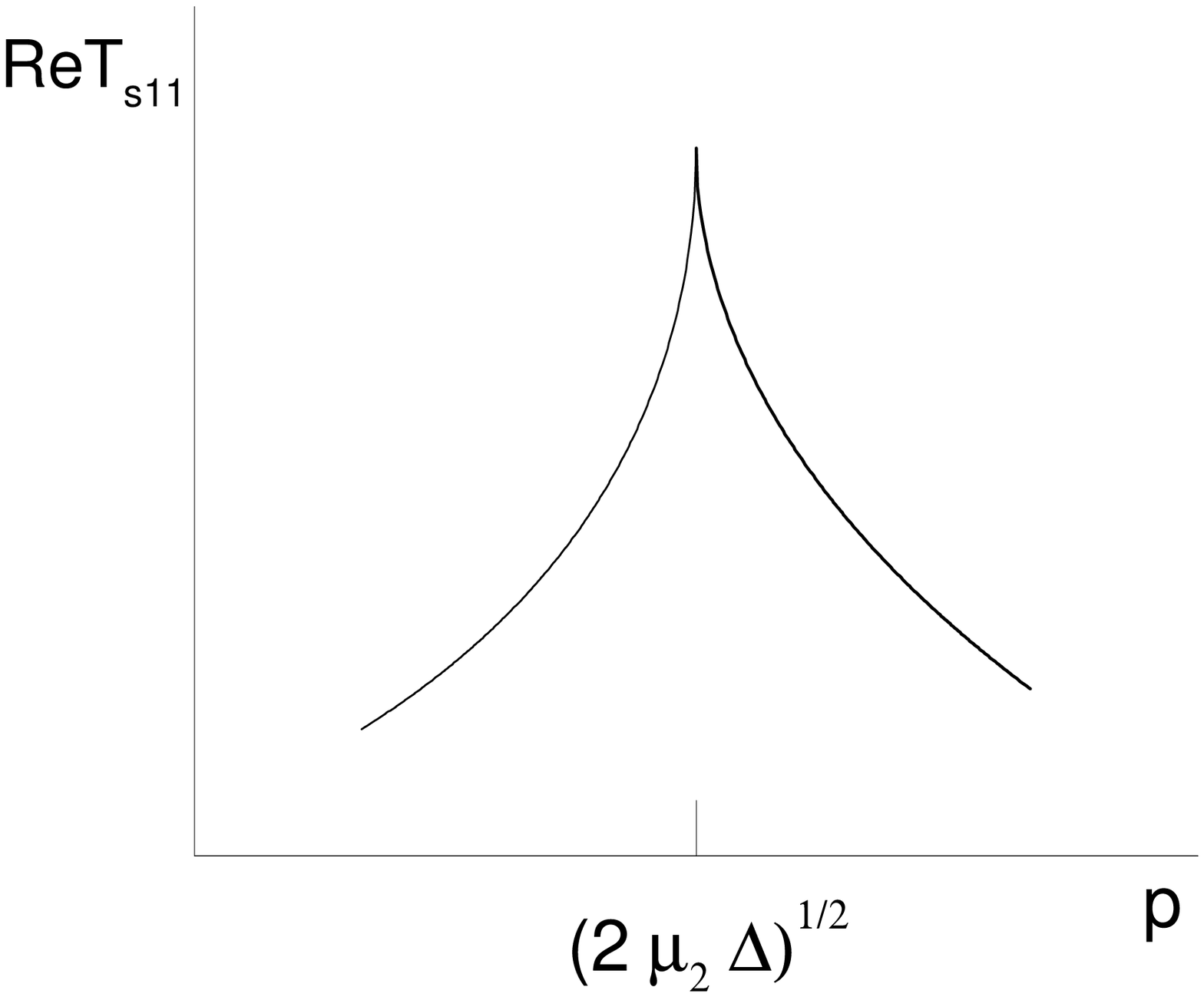}
\label{fig1}
\end{minipage}
\hfill
\begin{minipage}[l]{1\columnwidth}
\centering
\includegraphics[height=7cm,width=9cm]{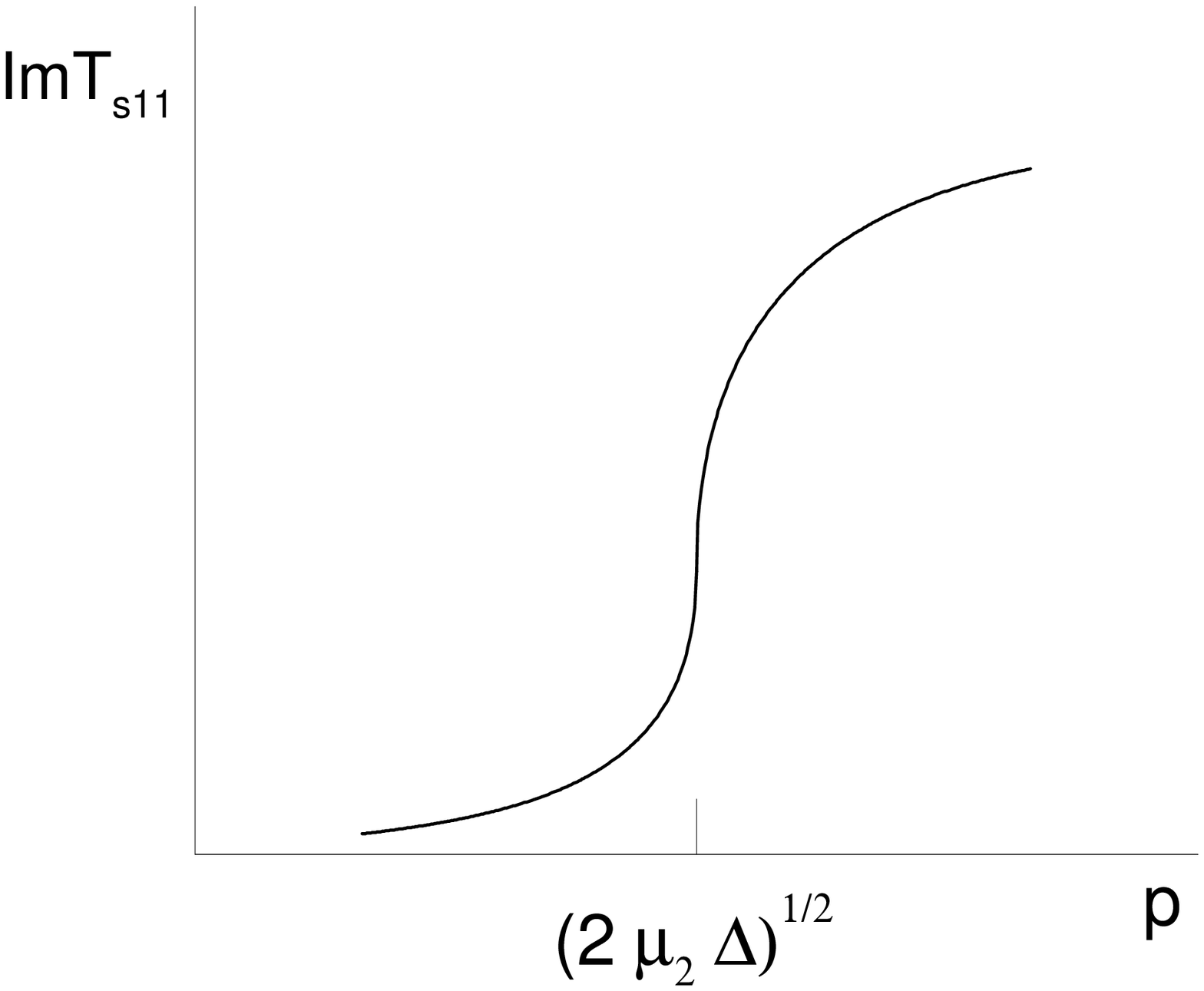}
\caption{Cusp and rounded-step behavior of the real and imaginary parts
of $T_{s11}$ near the 
threshold for channel ${\rm II}$.}
\label{fig2}
\end{minipage}
\hfill
\end{figure}

As the energy $E$ increases and approaches 
the kinematical threshold for channel ${\rm II}$ from below,
the $i\sqrt{p^2-2\mu_2\Delta}$'s in Eq.~(\ref{TT})
---which are real--- decrease, and 
above the threshold they acquire an imaginary part.
At $p_{\rm t}=\sqrt{2\mu_2 \Delta}$,
the amplitude
for channel ${\rm I}$ ---given by $T_{s11}$--- 
is continuous, but its derivatives are not.
The derivatives at threshold 
explode as 
$1/\sqrt{-(p-p_{\rm t})}$ from below, and as $1/\sqrt{p-p_{\rm t}}$
from above.
The sign of the derivatives is governed by 
$$\lambda= \left(\frac{1}{a_{11}}-\frac{a_{22}}{a_{12}^2}\right)
           \frac{1}{\sqrt{2\mu_2 \Delta}}.$$
The sign of the derivative of the
real (imaginary) part of the amplitude is that of $\lambda^2-1$ 
($-\lambda$) below,
and $\lambda$ ($\lambda^2-1$) above threshold.
As a consequence, the amplitude exhibits a cusp 
and a rounded step \cite{L}. 
Fig.~\ref{fig2} illustrates this behavior for 
the particular combination
$\lambda<0$ and $\lambda^2-1>0$.

\section{Optical potential and EFT \label{optical} }

Optical potentials ---potentials with real and imaginary
parts--- are usually introduced to describe scattering in which
the flux in the sector explicitly considered is not conserved.
This occurs when there are open channels that are not explicitly
described by the theory. A well-known example in hadronic physics
is $p\,\bar{p}$ annihilation at low
energies \cite{opticalNN,NantiN}. 
It is clearly of interest to develop an
effective-field-theory treatment for such cases.  Presumably, the
generic logic underlying EFT's power counting applies.  
The subtle issue concerns whether one can integrate out a
channel that is separated by a large mass scale 
but is open, that is, represents asymptotic states
that are accessible even at low energy.
Certainly, if one does this one can no longer impose unitarity. 
Thus it is
plausible that the form of an EFT in such a regime would be that
of a usual EFT with usual power counting but with complex
coefficients
\footnote{Such an EFT has been used, for example, to describe the annihilation
decay rates of positronium and heavy quarkonium \cite{quark},
and three-atom recombination into an atom and a deep dimer \cite{dimer}}.  
We also expect that the result would be an expansion
analogous to the effective range expansion, with complex parameters; 
indeed, such expansion already exists \cite{nnbar}. 
If this can be established, then the EFT approach
can be usefully extended to a wide array of new systems.

One way to test this idea is in the context of a solvable model.
Consider, for example, the coupled-channel scattering
discussed in section~\ref{cch} but with negative $\Delta$.
Assume furthermore that $\sqrt{\mu_2 |\Delta|}$ is large compared to
the other low scales in the problem ($1/a_{ij}$ and $p$) but
remains small compared to the underlying scale
$M$, which justifies the treatment in the previous section.
In this case we have a simple example of such an EFT and we
can test how it works.

To see how the coupled-channel calculation goes with these
kinematics, we only need to slightly modify the treatment. In
section~\ref{cch} we discussed an EFT treatment of the
low-energy scattering with two open channels for which the
threshold $\Delta$ was positive so that the coupled channel was
above the initial channel ---channel ${\rm I}$. 
Yet, the treatment is valid also
when the second channel is below the initial channel. In fact the
form of the $T$-matrix in Eq.~(\ref{TT}) is the same with
just a single difference ---the threshold $\Delta$ is negative.
The renormalization is not affected by this change. Thus the
scattering in channel ${\rm I}$ is given by $T_{s11}$, Eq.~(\ref{TT}).
While channel ${\rm II}$ is now open for any non-zero initial
energy, $T_{s11}$ can still be expanded in powers of 
$E/|\Delta|$
when $E < |\Delta|$. The first three terms in such an expansion are, again,
\begin{widetext}
\begin{equation}
{1\over T_{s11}} ={\mu_1 \over 2\pi} \left\{-{1\over a_{\rm eff}}
+ {1\over 2} r_{\rm eff} p^2 -i\, p +{\cal O}
\left({p^4\over a_{12}^2 (\mu_2 |\Delta|)^{5/2}}\right)\right\} \, , 
\label{efr2}
\end{equation}
\end{widetext}
but now the expansion parameters are given by
\begin{eqnarray}
a_{\rm eff} &\equiv &  {a_{11} a_{12}^{2} (1+ia_{22}\sqrt{2 \mu_{2} |\Delta|})
\over a_{12}^{2} (1+ia_{22}\sqrt{2 \mu_{2} |\Delta|})-a_{11} a_{22}}
 \; , \nonumber \\
r_{\rm eff}  &\equiv &  -
{i\over \sqrt{2 \mu_{2} |\Delta|}} 
\left({a_{22}\over a_{12} (1+ia_{22}\sqrt{2 \mu_{2} |\Delta|})}\right)^2. 
\label{complex}
\end{eqnarray}
The expansion in Eq.~(\ref{efr2}) has the form of the effective
range expansion with complex coefficients that reflect the
scales in the problem.
While the effective scattering length is approximately
real and large, $a_{\rm eff}\simeq a_{11}$,
the effective range is approximately imaginary and small,
$r_{\rm eff}\simeq i/(a_{12}^2 (2 \mu_{2} |\Delta|)^{3/2})$. 

We expect, therefore, that the EFT developed here can be
generalized to describe other problems with open channels well
below the scale explicitly included in the EFT.  It remains an
open question how relevant such a treatment would be for $p\,
\bar{p}$ scattering at low energy. The obvious complication is the
presence of many open channels containing mesons in the final
states. One, however, can exploit the scale separation between
the elastic and charge-exchange channels (with only nucleons and
anti-nucleons in the final states), and the channels containing
mesons. As far as the EFT description of $p\,\bar{p} \rightarrow
p\,\bar{p}$ and $p\,\bar{p} \rightarrow n\,\bar{n}$ channels 
is concerned, the
rest of the channels can perhaps be described by a single effective
channel separated by a threshold $\Delta$.

\section{Summary}

We have constructed an effective field theory for the low-energy
coupled-channel scattering with short-range forces. The EFT is
valid for the scattering with an unnaturally-large cross section and
an unnaturally-small threshold between channels. At leading order, the
$s$-wave $T$-matrix in the two-channel case 
is given in Eq.~(\ref{TT}) and is determined
by three momentum-independent contact interactions shown in
Eq.~(\ref{Lint}). This $T$-matrix is expressed in terms of three
renormalized quantities ---$a_{11}$, $a_{22}$ and $a_{12}$--- that
are analogous to the scattering length in the single-channel case. 
The treatment here can be readily extended to include
higher orders and more channels \footnote{After the present
paper was completed, the authors became aware of the discussions of
coupled-channel scattering in Ref. \cite{BH}
---where the same Eqs.~(\ref{cii}) and (\ref{TT}) were obtained---
and in Ref. \cite{delta} ---where the $\Delta\Delta$ channel was 
explicitly calculated in $S$-wave $NN$ scattering. 
Our analysis is, however, considerably more detailed.
We thank B.R. Holstein for communication on the subject and
encouragement.}.

In addition, we have also discussed the case where the second
channel is energetically below the initial channel. In this case
the scattering in the initial channel can be described by a
generalized effective range expansion with complex coefficients,
Eqs.~(\ref{efr2}) and (\ref{complex}).

\section*{Acknowledgments}
The authors
would like to thank M. Alberg, B. El-Bennich and R.G.E. Timmermans for useful
discussions about optical potentials and the $\bar N N$ system.
The authors acknowledge the hospitality of the Institute for
Nuclear Theory at the University of Washington, Seattle, where most
of this research was conducted during the Fall 2003 INT Program on
``Theories of Nuclear Forces and Nuclear Systems''. 
B.A.G. gratefully
acknowledges the hospitality of the staff and members of the
Theory Group for Quarks, Hadrons and Nuclei at the University of
Maryland, College Park, where part of the work was done.
U.v.K. thanks RIKEN, Brookhaven National Laboratory and 
the U.S. Department of Energy [DE-AC02-98CH10886] for providing the facilities
essential for the completion of this work.
This work was supported in part by
the U.S. Department of
Energy under grants DE-FG02-93ER-40762 (T.D.C.)
and DE-FG03-01ER-41196 (B.A.G., U.v.K.),
and by the Alfred P. Sloan Foundation (U.v.K.).

\end{document}